\documentclass[%
 reprint,
 amsmath,amssymb,
 aps,
]{revtex4-2}

\usepackage{lineno}
\usepackage{graphicx}
\usepackage{dcolumn}
\usepackage{amssymb}
\usepackage{amsfonts}
\usepackage{fancyhdr}
\usepackage{epsfig}
\usepackage{amsmath}
\usepackage{amssymb}
\usepackage{bm}
\usepackage{bbm}
\usepackage{ulem} 
\usepackage{mathrsfs}

\newcommand\T{\Theta_{T}}
\newcommand\vx{\vec{x}}

\newcommand\tr{\text{Tr}}

\def\ket#1{\left| #1 \right\rangle}

\begin{document}

\preprint{APS/123-QED}

\title{CRT gauge symmetry in two-sheet de Sitter universes}
\author{Ji-Yu Cheng}
\email{chengjiyu@mail.ustc.edu.cn}
\affiliation{Deep Space Exploration Laboratory/School of Physical Sciences,
University of Science and Technology of China, Hefei, Anhui 230026, China \label{addr1}
\\
CAS Key Laboratory for Researches in Galaxies and Cosmology/Department of Astronomy,
School of Astronomy and Space Science, University of Science and Technology of China, Hefei, Anhui 230026, China \label{addr2}}
\date{November 17, 2024}

\begin{abstract}
We show how to understand CRT symmetry as gauge symmetry in holographic de Sitter universe involving a pair of mirror universes, in which frame there are two times going in two opposite temporal directions each. The CRT symmetry is global with respect to the big bang for these two-sheet universes. In this construction the presence of an observer is equivalent to taking a CRT gauge, thus breaks the global CRT symmetry. In de Sitter space the observer need not to carry a clock, as the clock is equivalent to a way telling the temporal order of events. The role of a clock can be replaced by redshift, like in FLRW cosmology, points to the adaptability of measurement in cosmological contexts. This concept drives us to write gauge invariant formulas, i.e. correlation function $[\mathcal{O}(t-w),\mathcal{O}(-t-w)]=0$ identically for $w$ the time from the observer to big bang. In this spirit, the gauge invariant formula overlaps with the fact that the Hilbert space is real in holographic de Sitter universe. The observer is highly entangled with the rest of the universe in a thermofield double state, thus contributes to no extra degrees of freedom. We show that the time $w$ to the big bang is not real singularity, but an unavoidable focus point along the timelike geodesic congruence of the observer. Our result highlights the role of observer in understanding quantum gravity.
\end{abstract}

\maketitle
\textit{Introduction.}---It's widely accepted that there is no global symmetry in theory of quantum gravity, including no global discrete symmetries. The CRT symmetry (C for particle-antiparticle exchanging, R for spatial coordinates reflection $x^i\rightarrow-x^i$, and T for time reversal $t\rightarrow-t$) are symmetries in relativistic (or equivalently Lorentz invariant, e.g., scale invariance in FLRW universe) quantum field theories \cite{harlow2021symmetryinqg,goodhew2024cosmologicalcrt}, which should also be gauge symmetries in quantum gravity. In Ref.\cite{harlow2023crt} the authors show how to understand CRT symmetries as gauge symmetries in holographic theories \cite{susskind2021dsholography,susskind1995holography,bousso1999covariant,bousso1999holography}, in which frame the global symmetries of the boundary theory correspond to gauge symmetries of the bulk. An intuitive explanation is, the spacetime admits M\"obius-like topological closed geodesic, and part of the geodesic is behind the horizon with respect to the observer, such that the causality on the boundary theory is preserved. This construction is topological and is valid for fields $\phi(x)$ in any spacetime points. In this letter we extend the CRT gauge symmetry in a closed de Sitter (dS) universe, demonstrate that taking the CRT gauge is equivalent to the presence of an observer. The procedure involves the construction of two-sheet universes \cite{boyle2018cptuniverse,boyle2021twosheetuniverse,boyle2022bigbangmirror}, where the CRT symmetry results in two mirror universes start from the singularity (big bang) and with opposite time direction $-t$. The two-sheet universes is supported by experimental evidence \cite{anchordoqui2018crtdarkmatter}. We assume holographic theory for closed de Sitter universe, where the observers are fluctuations in it \cite{susskind2021dsholography,susskind2023dsparadox,clpw2023dsalgebra,chen2024clock}.

The reason for considering an observer inside dS space is that, the work of \cite{clpw2023dsalgebra} showed that the dS quantum gravity under semi-classical condition can be described by Type-II von Neumann algebra. The algebra tells us the whole dS universe has a maximum entropy, which contains an effective renormalizable trace, thus was understandable, with the necessity of introducing an observer inside \cite{sorce2023notes}. The presence of an observer is usually recognized as a clock-carrier \cite{clpw2023dsalgebra,witten2024observer,witten2024backgroundindependent}, which means it must carry at least one extra degrees of freedom (DoF). Philosophically speaking, a clock is just a way to tell the temporal order of events, thus in principle we can use many things to play the role of a clock, such as redshift, inflation potential, black hole radiation, etc \cite{chen2024clock}. In dS space an observer can be a fluctuation which will not create any back-reaction to background such that to recover the $G_N\rightarrow0$ limit to introduce the observer, and that's why the defined observer accesses a background independent algebra \cite{clpw2023dsalgebra,witten2024backgroundindependent}. This property is actually irrelevant to the observer, but because the dS space records its history over itself, which makes sure whenever an observer births, he is able to trace back the history of the dS space until the coordinate system fails \cite{hawking1966singularities}.

\textit{Preliminaries.}---In this letter we basically demonstrate a way to understand the time reversal symmetry as a gauge symmetry in a closed universe, also shows that taking CRT gauge is equivalent to choosing an observer. It's usually understood that the time reversal (T-reversal) is a backward going along the known history, where the second law of thermodynamics cannot be valid along the backward going history. We stress that the word history means the recorded facts, with definite measurements on physical quantities, at the level of both quantum mechanically and classically. This is the fundamental concept which makes the backward going time fields are not the replaying the going back history, but an individual varying object regardless of the real history of forward going fields. To show the difference between history and backward going fields, just imagine the black hole area decreases classically as time $t$ flows to $-t$ \cite{bekenstein1974gen2ndlawbh}. To formulate the CRT symmetry it's necessary and natural by introducing a pair of mirror universe \cite{boyle2018cptuniverse}, where the forward and backward going times lives in different part of spacetime manifolds. The positive and negative time directions are the time direction with respect to the big bang, where the mirror sits. We prove this construction solves the paradox that the dS holographic theory has real Hilbert space \cite{harlow2023crt,susskind2023dsparadox}, and the holographic principle is valid on static patch, as there is no extra DoF involves due to the entanglement.

In this letter we work in de Sitter holographic theory. To make it self-consistent we introduce three principles:

(i). The observer presents as fluctuation in dS holographic theory, which can have a comparably (with the lifetime of de Sitter space) short or long lifetime \cite{clpw2023dsalgebra}.

(ii). The observer is in maximal entanglement state with the rest DoF in the closed universe \cite{susskind2023dsparadox} as well as with the observer in the mirror universe with opposite temporal direction \cite{boyle2018cptuniverse,boyle2021twosheetuniverse,boyle2022bigbangmirror}.

(iii). The clock is a classical way to tell the temporal order of events, thus the concept of time can be replaced by other quantities, especially those who show the capability of distinguishing the temporal order in history \cite{chen2024clock}. Thus, in de Sitter space we do not require the observer to carry a clock.

Before entering the main part let's make some remarks. The first principle is necessary to understand the dS quantum gravity, at least in semi-classical limit \cite{clpw2023dsalgebra}. We will show the Hermitian property of operators is related to the CRT symmetry, explicitly, we show that it's CRT symmetry which restricts the Hamiltonian and momentum of the observer must have real eigenvalues, and that's why we understand the dS holographic theory has real Hilbert space. We demonstrate the observer is the recorder of the history, thus it can be used as the clock, like inflation potential \cite{chen2024clock}. What's more, the principles (ii) and (iii) are the resolution of a paradox raised in Ref.\cite{harlow2023crt} that the introduced clock carried by an observer would violate the principle of holography, and results in breaking CRT symmetry. We argue that we can always use the recorded history as the clock when we choose the coordinates origin of the birth of fluctuation, and it's not the interactions between the clock and the background but the choosing of observer breaks the CRT symmetry in closed dS universe. We show that even the origin of the coordinates, like a singularity, is due to the observer effect. This is the key point to understand CRT gauge symmetry.

\textit{Observer in holographic de Sitter two-sheet universes.}---In de Sitter space the observer is a fluctuation consists of energy levels sitting in the middle of static patch, where a holographic dS quantum gravity must contain the DoF of the observer and take it as a whole. In this letter we always assume a holographic theory, which means the DoF of the observer are contained in total DoF on the co-dimensional 2 hologram surface, the static patch \cite{shaghoulian2022dsentanglement}. We assume the de Sitter space has a starting point/big bang/singularity as the mirror and has a mirrored twin universe, as indicated in \cite{boyle2018cptuniverse,boyle2021twosheetuniverse,boyle2022bigbangmirror}, in this section we assume there exists a fluctuation which was born very close to the big bang, this is an observer started to record the history from the approximately beginning of the two universes.

We start with considering a simple fluctuation with Hamiltonian with flat spectrum $H_+$, which is a comparably small energy so that it has no back reaction to spacetime background. In Ref.\cite{harlow2023crt} the authors pointed out the CRT symmetric state are M\"obius pseudostates, which means the operators are not Hermitian in such a Hilbert space. Thus, we do not require the Hamiltonian is Hermitian, but we show that it will become strictly Hermitian operator with real eigenvalues in the end. We write the Hamiltonian of the fluctuation as
\begin{equation}
H_+=A+iB,
\end{equation}
where $A$ and $B$ are real. Such Hamiltonian gives solution of Schrodinger equation $\psi(\tau)\propto e^{-i(A+iB)\tau}$, where $\tau$ is the conjugate time. The probabilistic interpretation of wave function is the basic principle of quantum theory, and we assume it applies in quantum gravity. The probability distribution for this wave function is $P=|\psi|^2\propto e^{2B\tau}$. We haven't defined the arrow of time yet, but in principle, such a fluctuation should have a two directions spreading. 

One may find $P$ is as the same as the other one, generated by taking a new Hamiltonian $H_-=A-iB$, but with the conjugate time $-\tau$. The construction is not a coincidence, because the Hamiltonian $H_-$ is the T-reversal of $H_+$. It's easy to check $\T H_+\T^{-1}=H_-$, where for T-reversal operator we have $\T\ket{\psi(\tau)}=\ket{\psi(-\tau)}$ and $\T^2=1$, for scalar fields \cite{goodhew2024cosmologicalcrt}. Therefore, the fluctuation gives rise to a generic combined system of two wave packets with two dynamically equivalent probability distributions but opposite conjugate times, one for going forward along the direction $+\tau$ and the other backward along $-\tau$. We write the Hamiltonian for the whole system as $H=\text{diag}(H_+,H_-)$. Under this construction, the two wave packets have the same origin of the fluctuation, in principle the two are entangled. Thus, we can write the generic Hamiltonian for this combined system as
\begin{equation}
H=\begin{pmatrix}
A+iB&M\\
M&A-iB
\end{pmatrix},
\end{equation}
where $M$ is the coupled constant of two wave packets. The Hamiltonian $H$ is T-invariant, and with two eigenvalues
\begin{equation}
E_\pm=A\pm\sqrt{M^2-B^2}.
\end{equation}
When the two systems are weakly coupled, which means $M^2<B^2$, the two eigenvalues are inequivalent, and one increases with time while the other decreases. This is not our case, because it breaks the T-reversal invariance, which means the fluctuation prefers a weight on one direction of time, which is unacceptable. Our assumption is, the T-reversal invariant system reveals two entangled states, which means $M^2>B^2$, $M^2\gg A^2$. This gives us a pair of eigenvalues $E_\pm\approx\pm M$, and the Hamiltonian is in form of
\begin{equation}
H=\begin{pmatrix}
M&0\\
0&-M
\end{pmatrix}.
\end{equation}
This state is actually our only choice, as the other forms of Hamiltonian satisfying T-reversal symmetry give different energy eigenvalues, which breaks the assumption of zero Hamiltonian for quantum gravity theory. The complex Hamiltonian assumption is not necessary, as if we set $H_+=A$ then the result is the same. The only difference is the probability distribution are all zero, but the corresponding Hamiltonian are still T-reversal. We can write the state in energy representation of observer:
\begin{equation}
\ket{\Psi_{obs}}=\sum_i\ket{E_i}_+\otimes\ket{-E_i}_-.
\end{equation}
We label the universe and quantities with and conjugates to positive temporal direction as $+$, and negative with $-$. There is no prohibition mechanism restricts the observer to choose a preferred time direction, thus the general state of observer is a combined state of two wave packets entangled with each other.

In a closed universe the net probability distribution should be identically zero. The observer contributes to $P\propto e^{-2B\tau}$, the other part of the universe must contributes to another so that the net vanishes. This means the rest of the universe must absorb the distribution. This process is just a T-reversal of $P(\tau)\rightarrow P(-\tau)$, which is equivalent to a corresponding wave function $\phi(\tau)\propto e^{i(A-iB)\tau}$ with Hamiltonian $H_-$. In this case, we reformulate the situation that the observer and the rest DoF are entangled with each other, so that the total energy is identically zero \cite{susskind2023dsparadox}. The state of this universe is
\begin{equation}
\ket{\Psi_{tot}}=\frac{1}{\sqrt{2}}\sum_{j=+,-}e^{i\theta_j}\ket{\tau_j}\otimes\ket{\Phi(\tau_j)},
\end{equation}
where $\theta_j$ are phases of two states $\ket{\tau_j}$, and $\ket{\Phi(\tau_i)}$ represents the rest DoF in the universe. The general state consists of a forward going and a mirrored state on backward going time direction, thus the above state is T-reversal invariant. In this case, the T-reversal gives two opposite energy eigenvalues with total energy zero, which corresponds to quantum reference frame (QRF) construction of dS holography. It's believed that such a construction of Hilbert space with respect to the observer who carries a clock would contribute to an additional DoF in the holographic theory, but we will argue that the holographic principle is preserved and it raises no contribution in de Sitter space. The key point is, the observer is entangled with the rest of the universe, thus the total degrees of freedom is unchanged, no matter how many discrete symmetries the observer gauges. These states are all in QRF states, which makes sure the commutator $C(\tau)=[\mathcal{O}(\tau),\mathcal{O}(-\tau)]=-C(-\tau)=0$ identically for any operator $\mathcal{O}$, as one can check the symmetric operators commute for QRF states, and solves the paradox of non-zero imaginary part of correlation function holographic dS theory \cite{harlow2023crt,susskind2021dsholography}. This statement comes from the expansion result that the dynamical quantities has CRT symmetry \cite{boyle2018cptuniverse}, thus they are all in QRF states, for example $\ket{\phi}=\sum\ket{p_i}\otimes\ket{-p_i}$, etc. That's why these pairs always commute. In this case, the observer and the rest DoF, and the observers, and the two universes, are all in a thermofield double state, with the form of
\begin{equation}
\ket{TFD}=\frac{1}{\sqrt{Z}}\sum_i e^{-\beta E_i}\ket{\bar{i}}\otimes\ket{i},
\end{equation}
or equivalently write density matrix $\rho_T=e^{-\pi K}\T$ as the two T-symmetric states have an angular difference of $\pi$ in analytic continuation of Euclidean spacetime, where in this case $K$ is the boost/modular Hamiltonian. This result is in analog with the M\"obius pseudostate for T-symmetry $\rho=e^{-\pi K}\T$ \cite{harlow2023crt}. The state $\ket{TFD}$ is a pure state, yet the observer and the rest DoF are in maximal entanglement. We have such a state if we set the birth of the observer at the big bang, and it's just an accident that the state is CRT symmetric. The more general situation is the birth of the observer can present at any time in dS space, it will eventually result in T-reversal gauge fixing.

Our result from two-sheet universes construction gives the thermofield double states, these are not the pseudostate like in \cite{harlow2023crt}, as the operators are Hermitian because of the entanglement. One can easily check that such state satisfies the CRT symmetry, i.e., CRT symmetry with density matrix $\rho_{CRT}=e^{-\pi K}\Theta_C\Theta_R\T$. This conclusion is natural, as the thermofield double state can be understood as a Euclidean path integral over two spacelike distinct CRT symmetric states, always labeled as $\ket{\phi_R}$ and $\ket{\phi_L}$ \cite{maldacena2003eternal}.

\textit{CRT gauge in de Sitter}---Our aim is to clarify what it means by taking a CRT gauge, how it breaks the CRT symmetry, and what is the gauge invariance in dS holography. It's suggested that the presence of an observer breaks the T-reversal symmetry, i.e. the correlation functions do not identically vanish any more, which give rise to a non-zero imaginary part $\langle[\mathcal{O}(\tau),\mathcal{O}(-\tau)]\rangle$ \cite{susskind2023dsparadox}, where $\tau$ is the conjugate time from the origin $\tau=0$ chosen by the observer. The observer origin can be shifted from the big bang. However, it raises a problem that, for example, the observer birth is in dS space at $\tau_0$ after big bang, then there exist a backward going observer wave packet in the temporal interval $\tau\in(0,\tau_0)$ with conjugate Hamiltonian $H_-$. It seems that in this temporal interval the total energy of the universe is $H=-2M$, where $E_-=-M$ for eigenvalue of backward going time direction wave packet. But remember the backward going observer goes in opposite time direction to the rest of DoF, thus the total energy of the universe should be $H=E_--E_-$ instead of $E_-+E_-$, thus the total energy is identical to zero \cite{maldacena2003eternal,maldacena2013coolhorizon}. In principle the backward going observer is not entangled with forward going dS rest DoF, but we know it's maximally entangled with forward observer, and the forward observer is maximally entangled with the forward dS, that's why we can still write the combined state in temporal interval $(0,\tau_0)$ as in a thermofield double state. In short, the conservation Hamiltonian of the total universe is due to the entanglement between observers and the rest.

The dS space is an expanding spacetime, like the period of inflation in our FLRW universe. Intuitively speaking, such an expanding $D$-dimensional spacetime should have a starting point, which is just a singularity where the volume form $v=\sqrt{-g}d^Dx$ collapses to zero. This singularity is where the two mirrored dS spaces meet, thus is always be recognized as the origin of spacetime $\tau,\vx=0$. The observer can find a way to calculate how much time it is after the origin, for example, by using Raychaudhuri equation \cite{hawking1966singularities}, as indicated by singularity theorem \cite{penrose1965singularity}. Another example is our FLRW universe with 13.8 billion year-old, supported by observational evidences. Although one observer can always find a way to calculate how much time it is after the singularity, he would find a different temporal interval to the singularity compared with another observer who has a shift $\Delta\tau$ on temporal direction. The two observers trace back the different history, the earlier observer would collect more information of the history than the later observer, this is one of the principle of cosmology.

No matter when the observer was born, he was able to calculate the proper time along the geodesic to the big bang. In general, the T-reversal symmetry is broken with respect to the observer, and since we have clarified what is the T-reversal gauge, our next job is to determine the meaning of CRT gauge invariance. We know that the CRT symmetry is valid in two-sheet universes with respect to the big bang, in this case we can always write out the gauge invariant correlation function
\begin{equation}
C_{gi}(\tau)=[\mathcal{O}(\tau-w),\mathcal{O}(-\tau-w)]=0,
\end{equation}
where $w$ is the proper time between the big bang and the observer at $\tau=0$, measured along the geodesic. The gauge invariant formula makes sure a dS holographic theory of quantum gravity is logical. The observer can always find when the big bang happens by calculating when the coordinate system fails to be valid \cite{hawking1966singularities}. We can always find such vanishing correlation function even after choosing different observers, with different $w$. In this case the CRT symmetry is preserved as a gauge symmetry with respect to the observer. The CRT gauge invariant correlation function is not an approximation but has precise physical meaning in dS holography \cite{susskind2023dsparadox}.


\textit{Big bang gauge.}---De Sitter space is an expanding spacetime at an exponential rate, which is the reason why an observer would see a causal horizon. In dS space we have global metric
\begin{equation}\label{dsmetric}
ds^2=-dt^2+a^2(t)\delta_{ij}dx^idx^j,\ \ \ \ a(t):=e^{t/l_{dS}},
\end{equation}
where $l_{dS}$ is the radius of its embedding manifold. The temporal length is equivalent to spacial distance in cosmology. We learn this because we can determine the distance in universe by redshift. In dS space the redshift is given by
\begin{equation}
z=e^{\frac{\Delta t}{l_{dS}}}-1,
\end{equation}
where $\Delta t=t_{now}-t_{then}$ is the time measured back from the observer to the earlier event. The horizon of the static patch is a hot event horizon. Thus the observer is able to trace back the history of the dS universe.

However, the observer cannot trace back to arbitrarily early history. Actually, the T-reversal gauge invariance in dS holography pushes us to re-think the meaning of CRT gauge. Although the presence of an observer gauges the CRT symmetry, the gauge invariant form seems to tell us that the dS space has a global CRT symmetry with respect to the big bang. From (\ref{dsmetric}) we find the Hubble parameter is given by $h\equiv\dot{a}/a=1/l_{dS}$, where the dot represents $d/dt$. In principle the observer can collect the information of the history of dS space and tell the time, like what we do in cosmology.

We assume the observer is at spacetime point $p$ on a codimension-1 spacelike hypersurface $S$, with a unit tangent vector $n$. The $D$-dimensional Raychaudhuri equation then reads:
\begin{equation}
\frac{d\theta}{d\lambda}=-\frac{1}{D-1}\theta^2-\sigma^{\mu\nu}\sigma_{\mu\nu}+\omega^{\mu\nu}\omega_{\mu\nu}-R_{\mu\nu}n^\mu n^\nu,
\end{equation}
where $\lambda$ is the affine parameter on the geodesic with tangent vector $n^\mu$. The global metric is equivalently describe an observer whose timelike geodesic parameterized by proper time $\tau=t$ is normal to the hypersurface $S$ passing through point $p$. In this case, we can replace the affine parameter $\lambda=\tau$ for geodesic, and the unit normal vector $n$. We can rewrite the Raychaudhuri equation as
\begin{equation}
R_{tt}=-\dot{\theta}-\frac{\theta^2}{D-1}-\tr[\sigma^2],
\end{equation}
assuming the universe has no rotation. Under the strong energy condition, we must have the inequality:
\begin{equation}
\dot{\theta}+\frac{\theta^2}{D-1}\leqslant0.
\end{equation}
Solving it we then conclude that any timelike geodesic normal to hypersurface $S$ that goes past will meet at least one focal point within $t-t_0=l_{dS}$ measured by proper time, with respect to the observer at $t$. It means the observer loses the ability to detect the big bang if it is born after the big bang greater than the proper time $l_{dS}$. Thus we are unable to find the real big bang as well as the real proper time $w$ from the observer to the big bang any more.

\textit{Discussions.}---The CRT symmetry is believed to be gauge symmetry in a theory of quantum gravity, in this letter we focus on answering two questions, 1). to formulate the CRT symmetry one must first indicate a spacetime coordinate origin, but which origin should we take? This question is answered by \cite{boyle2018cptuniverse}, that the universe is CRT symmetric with respect to the big bang. 2). if the CRT symmetry has a preferred origin, then how to make it a gauge? We demonstrate that the observer can never reaches the real big bang, as his timelike geodesics must end at a focus point after a proper time $t=l_{dS}$, no matter at what time the observer was born. This is the truth of the spacetime, that even the presence of the singularity is because of observer effects.
 
CRT symmetry is easier to understand in two-sheet universes construction, and a fluctuation gives a clearer definition of the symmetry of the wave packets of forward and reverse time evolution. We need to define a reference frame of time in order to define CRT symmetry, which corresponds to big bang's hypothesis. We explain why a two-sheet universes with a big bang still has a CRT gauge. The existence of CRT gauge is different from the coordinate selection invariance because the existence of CRT symmetry depends on the selection of big bang. We show that the observer cannot time the big bang, or even prove the existence of the big bang, because at $t=l_{dS}$ time the geodesic congruence always converges into a focus point.

To conclude, such results are generic means that the breaking of global CRT symmetry is not an isolated or particular occurrence but a fundamental aspect of how observers interact with their cosmological framework. It clarifies the interplay between fundamental physics and observational reality. In summary, our exploration into CRT symmetry as a gauge symmetry in holographic de Sitter universe offers profound insights into the nature of time, observation, and symmetry in cosmology. It aligns with the notion that our understanding of reality is deeply intertwined with the frameworks we apply and perspectives from which we observe the universe. This discussion opens avenues for further exploration in cosmology, quantum gravity, and the nature of time itself.

\textit{Acknowledgment}---The letter is especially dedicated to Jing-Qi Wang. The author would like to thank for her selfless encouragement and care during this work. It could not have been finished without her company and love. We are also grateful to Dong-Dong Zhang, Emmanuel N. Saridakis, and Yi-Fu Cai for helpful discussions and encouragements.

\bibliography{Reference}
\end{document}